\documentclass[aps,preprint,nofootinbib,superscriptaddress,prc]{revtex4}
\usepackage{booktabs}
\usepackage{amssymb}
\usepackage{indentfirst}
\usepackage{amsmath}
\usepackage{xcolor}
\usepackage{enumerate}
\usepackage{tabularx}
\usepackage{array}
\usepackage{makecell}
\usepackage{graphicx}
\usepackage{xcolor}
\usepackage{epsfig}
\usepackage[bookmarksnumbered,bookmarksopen,colorlinks,citecolor=blue,linkcolor=blue] {hyperref}
\begin{document}
	
	\date{\today}
	
\title{Extracting nuclear charge radii from binding energies: a single-parameter empirical formula with structural corrections}
	\author{Pengfei Ma}
	\affiliation{Department of Physics, Guangxi Normal University, Guilin 541004, People's Republic of China }

	\author{Minghui Hu}
	\affiliation{Department of Physics, Guangxi Normal University, Guilin 541004, People's Republic of China }
	
	\author{Kai Ren}
	\affiliation{Department of Physics, Guangxi Normal University, Guilin 541004, People's Republic of China }
	
	\author{Junlong Tian}
	\email{tianjl@gxnu.edu.cn}
	\affiliation{Department of Physics, Guangxi Normal University, Guilin 541004, People's Republic of China }
	\affiliation{Guangxi Key
		Laboratory of Nuclear Physics and Technology,  Guilin 541004, People's Republic of China}

\author{Cheng Li}
 \email{licheng@gxnu.edu.cn}
\affiliation{Department of Physics, Guangxi Normal University, Guilin 541004, People's Republic of China }
\affiliation{Guangxi Key
Laboratory of Nuclear Physics and Technology,  Guilin 541004, People's Republic of China}

\begin{abstract}
Nuclear binding energies and charge radii stem from the same underlying physics: saturation, isospin dependence, shell
structure, and deformation. Binding-energy data therefore provide a
natural constraint for charge-radius modeling. We propose a one-parameter charge-radius formula ($\mathrm{BECR}_\mathrm{1p}$) that combines
binding-energy correlations with local structural corrections. On a
curated set of 893 experimental charge radii, the macroscopic
$\mathrm{BECR}$ term alone reproduces the leading charge-radius
scale with a root-mean-square deviation (RMSD) of
$0.0345~\mathrm{fm}$; adding shell, odd--even, finite-size, and
deformation corrections further reduces the RMSD of
$\mathrm{BECR}_\mathrm{1p}$ to $0.0138~\mathrm{fm}$. An anisotropic kernel
ridge regression (AKRR) applied to the residuals further lowers the
leave-one-out cross-validation RMSD to about $0.0081~\mathrm{fm}$.
We use the formula to predict charge radii for 11205 nuclei across the nuclear chart.
\end{abstract}


\maketitle

\section{Introduction}
\label{sec:introduction}
The root-mean-square nuclear charge radius $\left(r_{\mathrm{ch}}\right)$
is a fundamental observable in nuclear physics, directly characterizing the spatial distribution of proton charge within a nucleus.
Its evolution along isotopic and isotonic chains offers
crucial insights into shell evolution, nuclear deformation, odd--even
staggering, and structural transitions far from the $\beta$-stability line~\cite{Angeli2013,Li2021,GarciaRuiz2016,
Miller2019,Gorges2019,deGroote2020,Koszorus2021}. The binding energy
per nucleon, $\epsilon = B/A$, quantifies the average binding strength
of a finite nucleus and simultaneously encodes information on
nuclear-matter saturation, surface effects, isospin asymmetry, shell
structure, and deformation. For a given nucleus, $r_{\mathrm{ch}}$ and
$\epsilon$ describe the same finite many-body system from
complementary perspectives: spatial extension and binding
strength. They may therefore be viewed as two intrinsically connected
facets of nuclear-structure information~\cite{Angeli1969,AngeliLombard1986,Angeli2015}.
A direct link between them would deepen our understanding of nuclear structure and provide a new empirical route to charge radii.

Most empirical radius formulae start from the volume scaling $R_{\mathrm{ch}} \sim A^{1/3}$ with corrections for isospin, shell, pairing, and deformation~\cite{WangLi2013,Sheng2015,Sun2014,Bao2020,Jiao2023}. They are simple and fast, but incomplete for local shell, deformation, and odd-even effects. Microscopic approaches, such as Hartree--Fock--Bogoliubov (HFB) theory, relativistic mean-field (RMF) models, and nuclear density-functional theory (DFT), are more firmly rooted in many-body theory, but their predictions depend sensitively on the adopted effective interaction or energy density functional and generally involve considerably more elaborate calculations~\cite{Bender2003,Goriely2016,Meng2006,Xia2018,Reinhard2017,An2020,An2024,Perera2021}.
In recent years,
data-driven and hybrid techniques, including kernel ridge regression,
radial basis functions, Gaussian processes, and neural networks, have
been increasingly employed to correct the residuals of empirical
formulae or microscopic models, yielding notable improvements in
numerical accuracy~\cite{Wu2020NN,Dong2022,Ma2020,MaZhang2022,
Tang2024NST,Li2026RBF,WuZhao2020}. The
physical interpretability and extrapolation reliability of these
methods still warrant careful scrutiny.
Experimentally, charge radii come from elastic electron scattering, muonic atoms, and laser spectroscopy~\cite{Angeli2013,Li2021,deVries1987,Sick2001,Campbell2016,Yang2023Review}. The database has grown with radioactive-beam facilities, but remains much smaller than nuclear mass data and is concentrated near the $\beta$-stability line; many neutron-rich and proton-rich regions lack radius measurements. Mass measurements, in contrast, cover a wider range. If structural information for charge radii can be extracted from $\epsilon$, it would provide a basis for radius estimates in data-scarce regions.

Building on this idea, we propose a charge-radius formula that uses $\epsilon$ as the explicit physical input, plus corrections for shell, pairing, finite-size, and deformation. This approach describes charge radii through binding-energy
correlations, rather than relying solely on the conventional $A^{1/3}$
scaling. After global optimization, the formula reduces to a one-parameter form (BECR$_{1p}$) with an RMSD of $0.0138$ fm on 893 experimental radii, reproducing shell kinks, odd-even staggering, and shape transitions.
An anisotropic kernel
ridge regression (AKRR) is subsequently applied to the residuals,
reducing the leave-one-out cross-validation (LOOCV) RMSD to
approximately $0.0081~\mathrm{fm}$. Machine learning is used here only as a residual-correction tool, not as a standalone predictor.

\section{Theoretical Framework}
\label{sec:theory}
\subsection{Definitions and reference data set}
The nuclear charge radius $r_{\mathrm{ch}}$ is defined as the
root-mean-square (rms) radius of the proton charge distribution
$\rho_{p}(\mathbf{r})$,
\begin{equation}
    r_{\mathrm{ch}}
    = \sqrt{\langle r^{2}\rangle}
    = \left[
        \frac{1}{Ze}\int r^{2}\,\rho_{p}(\mathbf{r})\,d^{3}r
      \right]^{1/2},
    \label{eq:rch_definition}
\end{equation}
with the normalization condition
$\int \rho_{p}(\mathbf{r})\,d^{3}r = Ze$. This quantity characterizes
the spatial extent of the proton charge distribution within a nucleus.
The binding energy per nucleon is denoted by $\epsilon = B/A$, where
$B$ is the (positive) nuclear binding energy. For a given nucleus,
$r_{\mathrm{ch}}$ and $\epsilon$ describe the same finite many-body
system from the spatial and energetic perspectives, respectively. The
average binding information encoded in $\epsilon$ can therefore serve
as a physically motivated input for extracting the nuclear
charge-radius scale.
\begin{figure}[htbp]
	\centering
	\includegraphics[width=\textwidth]{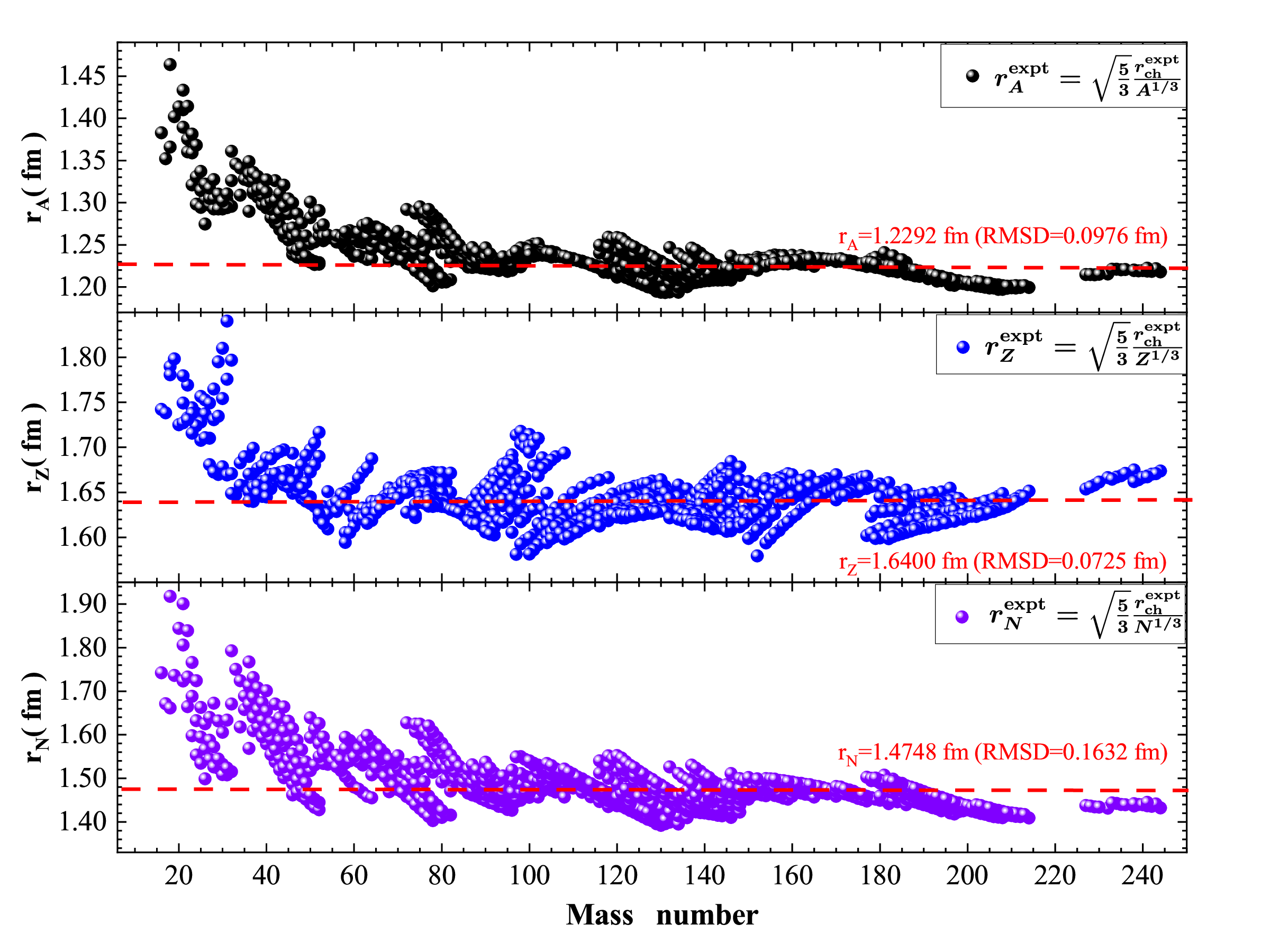}
	\caption{(Color online) Reduced experimental nuclear charge radii
		$ r_x = \sqrt{\frac{5}{3}}\frac{r_{\mathrm{ch}}^{\mathrm{expt}}}{x^{1/3}}$
		as functions of the mass number $A$, where $x=A,Z,N$.
		From top to bottom, the three panels correspond to reductions with
		$A^{1/3}$, $Z^{1/3}$, and $N^{1/3}$, respectively.
		The red dashed lines denote the fitted parameter values obtained for each reduced-radius representation.
	}
	\label{fig:reduced_radii_scaling}
\end{figure}

The experimental charge-radius data used in this work are compiled
from Refs.~\cite{Angeli2013,Li2021,Bai2025Sc}, comprising 1032 measured nuclear charge radii. To ensure the reliability of the
fitting procedure, several chains are excluded from the reference
data set. First, the Tb $(Z=65)$, Tm $(Z=69)$ and Lu $(Z=71)$ isotopic chains are removed, as they exhibit coherent
chainwise systematic offsets relative to neighboring nuclear regions
and to several theoretical models. The same exclusion was recently
adopted in a global charge-radius systematics study~\cite{Jiao2025}.
Second, nuclei whose absolute charge radii rely on \emph{calculated} rather than directly measured reference radii, including those in the Re $(Z=75)$, Po $(Z=84)$, Rn $(Z=86)$, Fr $(Z=87)$, Ra $(Z=88)$, and Cm $(Z=96)$ chains of the Angeli--Marinova
compilation~\cite{Angeli2013}, are also removed. After these selections, the reference data set contains 893 experimental charge radii; the excluded nuclei are not used in determining the global parameters, and the rationale for this treatment is discussed in Sec.~\ref{sec:results}.

To illustrate the limitation of a single geometrical scaling variable,
Fig.~\ref{fig:reduced_radii_scaling} shows the reduced experimental
nuclear charge radius $ r_x = \sqrt{\frac{5}{3}}\frac{r_{\mathrm{ch}}^{\mathrm{expt}}}{x^{1/3}}$, with $(x=A,Z,N)$,
plotted as a function of the mass number $A$. If the charge radius
were governed solely by a single volume-scaling variable, the
corresponding reduced radius would fluctuate around an approximately
constant value. However, for all three reductions based on $A^{1/3}$,
$Z^{1/3}$, and $N^{1/3}$, pronounced systematic drifts and local
fluctuations persist. This demonstrates that a purely geometrical
leading scale is insufficient to capture both the global trend and
the local deviations of nuclear charge radii, motivating the extraction of the charge-radius scale from $\epsilon$ instead.

\subsection{Macroscopic radius scale from the binding energy}
The macroscopic term extracts the leading charge-radius scale from $\epsilon$. Since $\epsilon$ and $r_{\mathrm{ch}}$ describe the same nucleus from complementary viewpoints, $\epsilon$ should carry spatial information.
We motivate the form through a volume–surface coupling picture. For a saturated nucleus, the neutron background gives a volume scale $R_N^3 \propto N$. In a finite potential, the tail length of a bound state scales as $\lambda_\epsilon \sim \hbar/\sqrt{2m_N\epsilon} \propto \epsilon^{-1/2}$. The surface diffuseness $a$ is related to this length, $a \sim \lambda_\epsilon$~\cite{AngeliLombard1986,Angeli2015}, so $a^2 \propto \epsilon^{-1}$. Combining these gives an effective length $L_{\mathrm{eff}} = (R_N^3 \lambda_\epsilon^2)^{1/5} \propto (N/\epsilon)^{1/5}$. This provides a phenomenological basis for the $1/5$ exponent.

However, this term grows with $N$ and overestimates the charge radius because $r_{\mathrm{ch}}$ probes protons, not total matter. We therefore subtract an isospin correction. The ratio $N/Z$ encodes the neutron excess, and we normalize it by $\sqrt{3}/2$—the invariant magnitude of the SU(2) isospin vector for a nucleon—yielding a negative correction $-\frac{\sqrt{3}}{2} N/Z$.

Based on the above considerations, the binding-energy-correlated
macroscopic charge-radius formula, denoted $\mathrm{BECR}^{\mathrm{mac}}$, reads
\begin{equation}
r_{\mathrm{ch}}^{\mathrm{mac}} =
\sqrt{\frac{3}{5}}
\left[
a_1 \left( \frac{N}{\epsilon/\mathrm{MeV}} \right)^{1/5}
- \frac{\sqrt{3}}{2} \frac{N}{Z}
- c_0
\right] \mathrm{fm},
\label{eq:macro}
\end{equation}
with $\epsilon$ from AME2020~\cite{AME2020}, and $a_1$, $c_0$ fitted. Fitting to the 893 experimental radii gives $a_1 = 5.6549$, $c_0 = 1.3723$, and an RMSD of $0.0345$ fm. To check these fixed choices, we also let the exponent and the isospin coefficient vary freely. The fit returns $\alpha = 0.2283$ and $\eta = 0.8689$, close to $1/5$ and $\sqrt{3}/2$, and the RMSD drops only slightly to $0.0336$ fm. This confirms that the physical constraints are well justified.

\subsection{Local structural corrections}

The macroscopic BECR term captures the leading charge-radius scale
through its dependence on the binding energy per nucleon $\epsilon$,
but it inevitably averages out local structural
effects~\cite{Angeli1979,Heyde2011,Otsuka2020,Geldhof2022}. To recover
these residual local correlations, we introduce physically motivated corrections to Eq.~(\ref{eq:macro}), guided by the following considerations:

 \textit{(i) Shell and odd--even corrections.} 
Local shell structure is described by a Casten-like
factor~\cite{Casten1985,Casten1987,CastenZamfir1993}, $e_{\mathrm{sh}}=N_p'N_n'/(N_p'+N_n')$, with $N_p'=|Z-Z_m|+1$ and $N_n'=|N-N_m|+1$. Here $Z_m$ and $N_m$ are the nearest proton and neutron magic numbers~\cite{Sheng2015}:
$Z_m = 2,\,6,\,14,\,28,\,50,\,82,\,114$ and
$N_m = 2,\,8,\,14,\,28,\,50,\,82,\,126,\,184$. 
Odd--even staggering is incorporated through the phenomenological
factor $e_{\mathrm{pair}}=+0.5$ (even-even), $-0.5$ (odd-odd), and $0$ (odd-$A$). These two contributions are combined into a single term,
$a_{2}(e_{\mathrm{sh}} + e_{\mathrm{pair}})/A^{5/6}$, which emulates the influence of shell gaps and pairing on the charge-radius scale, reproducing the characteristic kinks at magic numbers. The prefactor $A^{-5/6}$ suppresses local effects with increasing mass. Although this exponent is determined empirically, it may be understood as a combined scaling of surface diffuseness and the leading radius, where the diffuseness acquires additional $A$ dependence through its connection to $\epsilon$.

\textit{(ii) Finite-size and deformation corrections.} A $1/N$ term accounts for finite-size effects, particularly important in light nuclei. Deformation is incorporated through a multiplicative factor involving $\beta_2$ and $\beta_4$, taken from the WS3.3 mass table~\cite{WangWS2010}.

Combining these corrections with the macroscopic term, the full BECR formula reads
\begin{equation}
	r_{\mathrm{ch}}
	=\sqrt{\frac{3}{5}}\,
	\left[ 1+
	\frac{5r_{\beta}}{8\pi}
	\left(
	\beta_2^2+\beta_4^2
	\right)\right]
	\left[
	a_1
    \left(\frac{N}{\epsilon/\mathrm{MeV}}\right)^{1/5}
	-
	\frac{\sqrt{3}}{2}
	\frac{N}{Z}
	+
	a_2
	\frac{
		e_{\mathrm{sh}}+e_{\mathrm{pair}}
	}
	{A^{5/6}}
	+
	\frac{1}{N}-
	c_0
	\right]
	{\mathrm{fm}} .
	\label{eq:final_model}
\end{equation}
Here $\epsilon$, $\beta_2$, and $\beta_4$ enter as external
physical inputs, while $a_{1}$, $a_{2}$, $c_{0}$, and $r_{\beta}$
are treated as fitting parameters.
The optimized parameters are strongly correlated. We reduce them to one effective parameter $\kappa$:
$a_2\simeq\kappa$,
$a_1\simeq 3+\frac{2}{\kappa}$,
$r_{\beta}\simeq\kappa$,
$c_0\simeq\sqrt{2+\kappa}$.
A refit constrained by these relations gives $\kappa = 0.7190 \pm 0.0001$ and an RMSD of 0.0138 $\mathrm{fm}$, nearly identical to the unconstrained four-parameter result. These $\kappa$ relations reflect empirical correlations from the global fit rather than a microscopic derivation. With these substitutions, Eq.~(\ref{eq:final_model}) becomes the effective one-parameter formula $\mathrm{BECR}_\mathrm{1p}$.

\section{Results and Discussion}
\label{sec:results}
\subsection{Global accuracy and residual distribution}
\begin{table*}[htbp]
	\centering
	\small
	\setlength{\tabcolsep}{4pt}
	\renewcommand{\arraystretch}{1.35}
	\caption{Comparison of the RMSD (in fm) obtained by the
         $\mathrm{BECR}^{\mathrm{mac}}$ and $\mathrm{BECR}_\mathrm{1p}$
         formulae proposed in this work with two representative
         empirical formulae, evaluated on the same set of 893
         experimental charge radii. The WS3.3-R~\cite{WangLi2013} and Sheng~\cite{Sheng2015} results are obtained using their published parameter sets.
		}
	\label{tab:global_comparison}
	\resizebox{\textwidth}{!}{%
		\begin{tabular}{llll}
			\toprule[0.8pt]
			Formula & Expression & Parameters & RMSD (fm) \\
			\midrule[0.5pt]
			
			$\mathrm{BECR}^{\mathrm{mac}}$ &
			$\displaystyle
			r_{\mathrm{ch}}
			=\sqrt{\frac{3}{5}}\,
	     	\left[  a_1
			\left( \frac{N}{\epsilon/\mathrm{MeV}}  \right) ^{1/5}
			-
			\frac{\sqrt{3}}{2}\frac{N}{Z}
			-c_0 \right]  {\mathrm{fm}}
			$
			&
			\begin{tabular}[t]{@{}l@{}}
				$a_1=5.6549$\quad$c_0=1.3723$
			\end{tabular}
			&
			$0.0345$ \\
			
			\addlinespace[0.4em]
			
			$\mathrm{BECR}_\mathrm{1p}$&
            $\displaystyle
            \begin{aligned}
            	r_{\mathrm{ch}}
            	=&\sqrt{\frac{3}{5}}\,
               	\left[
            	1+\frac{5\kappa}{8\pi}
            	\left(\beta_2^2+\beta_4^2\right)
            	\right]
            	\left[
            	(3+\frac{2}{\kappa})
                \left(\frac{N}{\epsilon/\mathrm{MeV}}\right)^{1/5}
            	\right.\\
            	&\left.
            	-\frac{\sqrt{3}}{2}
            	\frac{N}{Z}
            	+\kappa
            	\frac{e_{\mathrm{sh}}+e_{\mathrm{pair}}}{A^{5/6}}
            	+\frac{1}{N}-\sqrt{2+\kappa}
            	\right]
            	\,\mathrm{fm}
            \end{aligned}
            $
			&
			\begin{tabular}[t]{@{}l@{}}
				$\kappa=0.7190$
			\end{tabular}
			&
			$0.0138$ \\
			
			\addlinespace[0.4em]
			
			WS3.3-R &
            $\displaystyle
            \begin{aligned}
            	r_{\mathrm{ch}}
            	=&\sqrt{\frac{3}{5}}\,
            	\biggl\{
            	\left[
            	1+\frac{5}{8\pi}
            	\left(\beta_2^2+\beta_4^2\right)
            	\right]
            	\left[
            	r_0A^{1/3}
            	+r_1A^{-2/3}
            	\right.\\
            	&\left.
            	-r_s I(1-I)
            	+r_d\frac{\Delta E}{A}
            	\right]
            	\biggr\}
            	\,\mathrm{fm}
            \end{aligned}
            $
			&
			\begin{tabular}[t]{@{}l@{}}
				$r_0=1.2261$\quad$r_1=2.8690$\\
				$r_s=1.0930$\quad$r_d=0.9917~\mathrm{fm/MeV}$
			\end{tabular}
			&
			$0.0170$ \\
			
			\addlinespace[0.4em]
			
			Sheng &
			$\displaystyle
			r_{\mathrm{ch}}
			=\sqrt{\frac{3}{5}}\,
			\left\lbrace r_0
			\left[
			1
			-a\frac{N-Z}{A}
			+b\frac{1}{A}
			+c\frac{P}{A}
			+d\frac{\delta}{A}
			\right]
			A^{1/3}\right\rbrace {\mathrm{fm}}
			$
			&
			\begin{tabular}[t]{@{}l@{}}
				$r_0=1.2293$   \quad$a=0.1576$\\
				$b=1.8327$\quad$c=0.4011$\quad$d=0.1308$
			\end{tabular}
			&
			$0.0247$ \\
			
			\bottomrule[0.8pt]
		\end{tabular}%
	}
\end{table*}
Table~\ref{tab:global_comparison} compares $\mathrm{BECR}^{\mathrm{mac}}$ and $\mathrm{BECR}_{1p}$ with WS3.3-R and Sheng on the same 893 experimental radii. $\mathrm{BECR}^{\mathrm{mac}}$, using only the $\epsilon$-based leading term and isospin correction, gives an RMSD of $0.0345~\mathrm{fm}$. It demonstrates that the average binding strength already captures the global trend. Adding shell, odd-even, finite-size, and deformation corrections brings $\mathrm{BECR}_{1p}$ to $0.0138~\mathrm{fm}$, better than WS3.3-R ($0.0170~\mathrm{fm}$)~\cite{WangLi2013} and Sheng ($0.0247~\mathrm{fm}$)~\cite{Sheng2015}. The macroscopic $\epsilon$-based term sets the effective leading scale; the local corrections then recover the structural details smoothed over in the macroscopic extraction.
\begin{figure*}[htbp]
	\centering
	\includegraphics[width=0.95\textwidth]{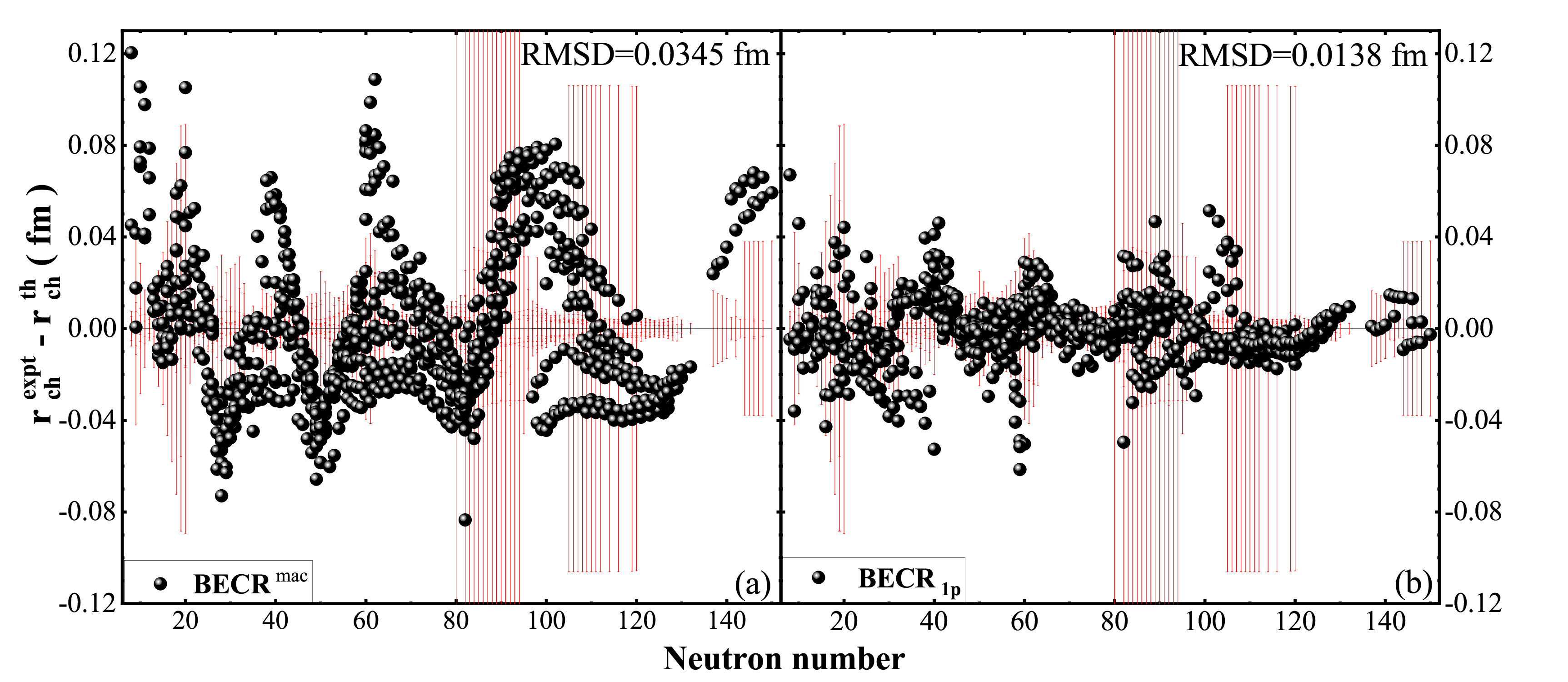}
	\caption{(Color online) Residuals $r_{\rm ch}^{\rm expt}-r_{\rm ch}^{\rm th}$ as functions of neutron number for the  $\mathrm{BECR}^{\mathrm{mac}}$ formula (a) and the $\mathrm{BECR}_\mathrm{1p}$ formula (b). Red capped vertical lines denote experimental uncertainties. The RMSDs are $0.0345~\mathrm{fm}$ and $0.0138~\mathrm{fm}$, respectively.
	}
	\label{fig:stageI_stageIV_residuals}
\end{figure*}

Figure~\ref{fig:stageI_stageIV_residuals} compares the residuals of $\mathrm{BECR}^{\mathrm{mac}}$ and $\mathrm{BECR}_{1p}$. The $\mathrm{BECR}^{\mathrm{mac}}$ residuals center near zero but show systematic structures in light nuclei, near shell closures, and in deformed regions: $\epsilon$ captures the global scale but misses local details. Adding shell, odd-even, finite-size, and deformation corrections compresses the residuals substantially; the RMSD drops from $0.0345$ to $0.0138~\mathrm{fm}$, and $89.4\%$ of the 893 nuclei fall within $\pm 0.02~\mathrm{fm}$. The improvement is systematic across the nuclear chart, not confined to particular mass regions. Residuals that remain may arise from higher-order shell effects, shape coexistence, rapid deformation changes, or experimental systematics.

\subsection{Performance along representative isotopic chains}
\begin{figure*}[htbp]
	\centering
	\includegraphics[trim=0.0cm 0.2cm 0.5cm 0.0cm,width=1.0\textwidth]{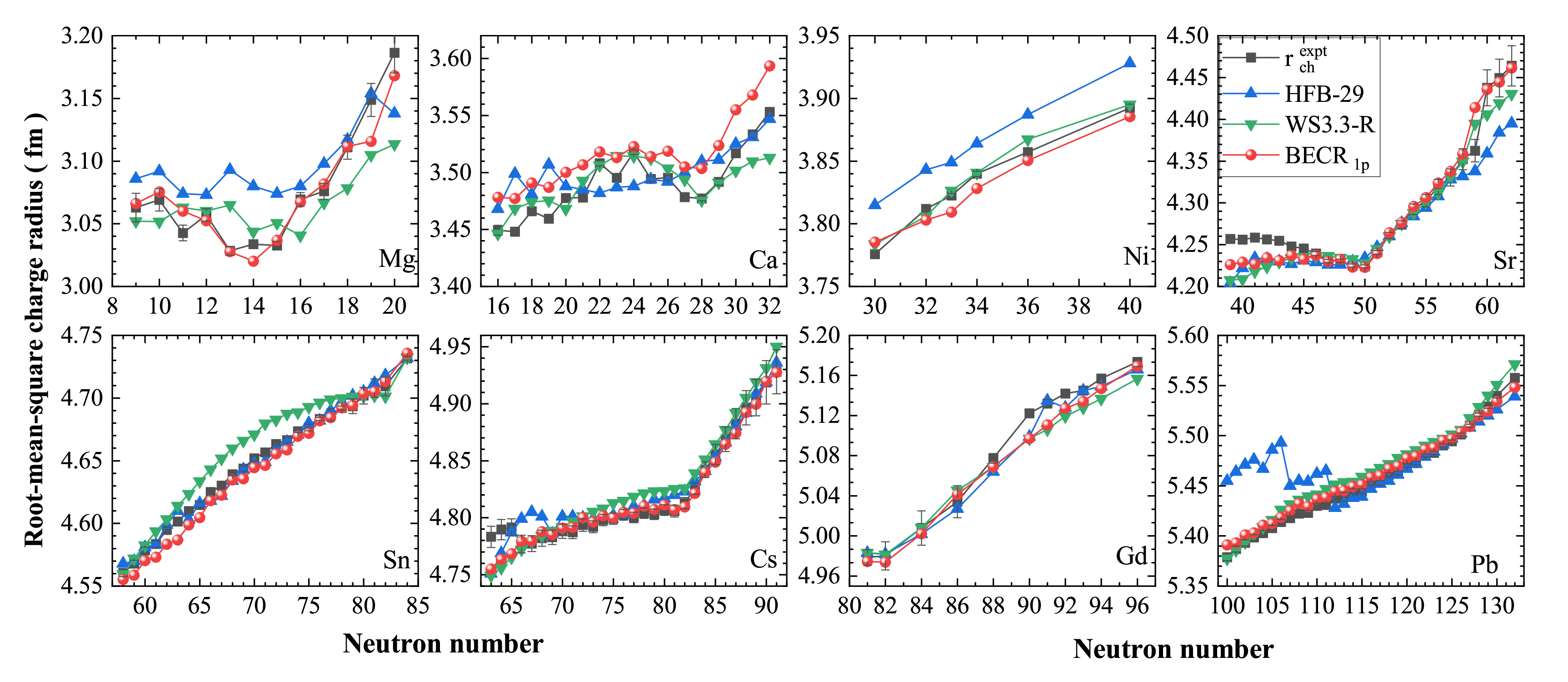}
	\caption{(Color online) Charge-radius evolution along the Mg, Ca, Ni, Sr, Sn, Cs, Gd, and Pb isotopic chains. Black squares: experimental data; red circles: $\mathrm{BECR}_\mathrm{1p}$; blue up-triangles: HFB-29; green down-triangles: WS3.3-R.
	}
	\label{fig:isotopes_benchmark}
\end{figure*}

To test $\mathrm{BECR}_{1p}$ on local structures, Fig.~\ref{fig:isotopes_benchmark} shows charge-radius evolution along eight chains (Mg, Ca, Ni, Sr, Sn, Cs, Gd, Pb), covering light to heavy nuclei, shell-closed and deformed regions. Predictions of HFB-29~\cite{Goriely2016} and WS3.3-R~\cite{WangLi2013} are included for comparison.
In light nuclei, surface and finite-size effects are strong. For Mg, HFB-29 overestimates and WS3.3-R deviates noticeably, while $\mathrm{BECR}_{1p}$ matches the data much better. For Ca, the $N=28$ shell closure and the odd-even staggering beyond it are well reproduced.
In medium-mass nuclei, Ni and Sn show smooth trends; $\mathrm{BECR}_{1p}$ produces no unphysical oscillations. The Sr chain has a sharp radius jump near $N\approx 60$ due to a shape transition, which the deformation correction captures reasonably.
For heavier Cs and Gd, where deformation is stronger, $\mathrm{BECR}_{1p}$ follows the data closely and even beats HFB-29 for some neutron-rich isotopes. The Pb kink at $N=126$ is also reproduced.
Overall, the $\epsilon$-based macroscopic scale plus the local corrections absorb most shell and deformation effects. $\mathrm{BECR}_{1p}$ thus reproduces not only the global RMSD, but also odd-even staggering, shape-transition jumps, and shell-closure kinks across the nuclear chart.

\subsection{Predictive power on unfitted chains}

\begin{figure*}[htbp]
	\centering
	\includegraphics[width=0.95\textwidth]{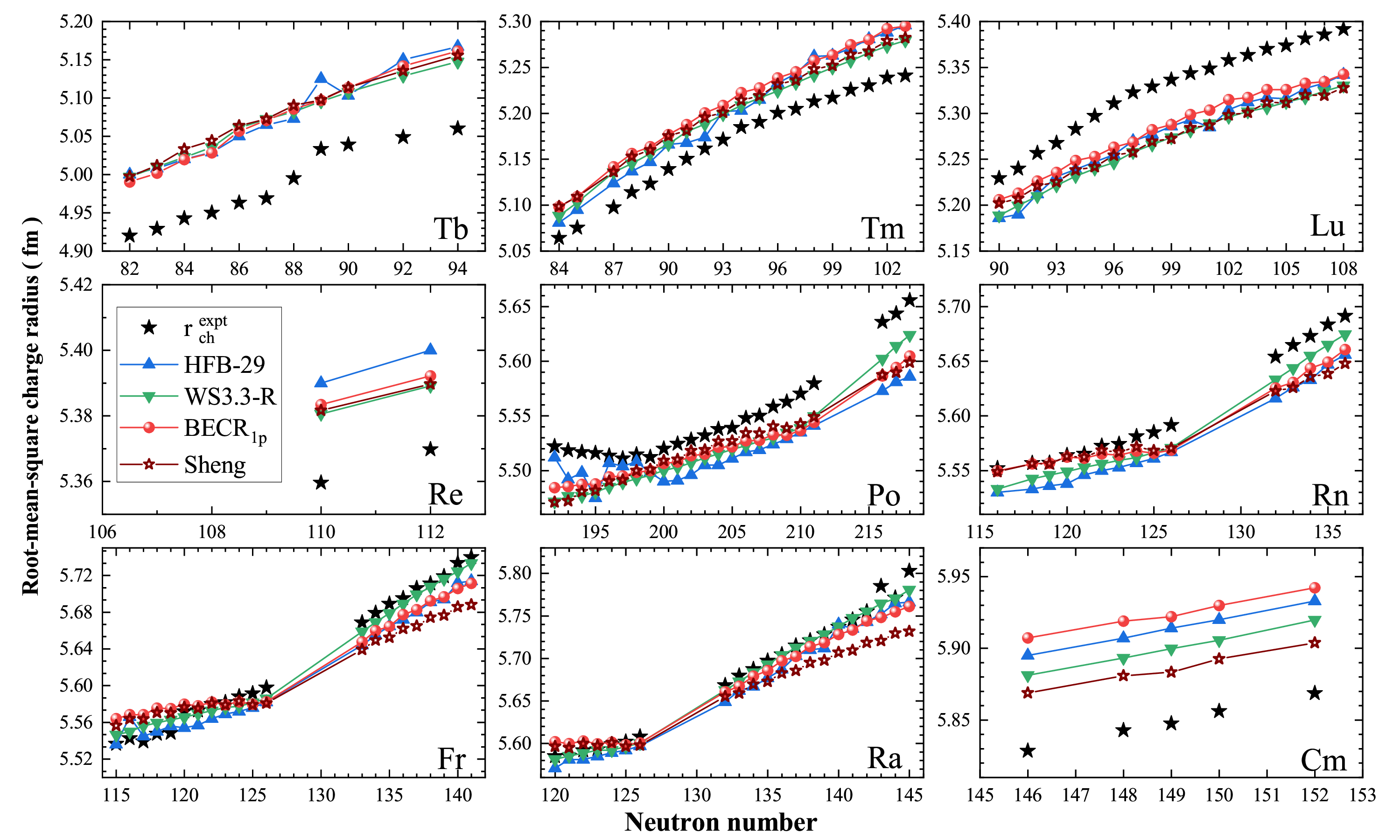}
	\caption{(Color online) Charge-radius evolution along nine isotopic chains excluded from the fit: Tb, Tm, Lu, Re, Po, Rn, Fr, Ra, and Cm. Black stars: experiment; red circles: $\mathrm{BECR}_\mathrm{1p}$; blue up-triangles: HFB-29; green down-triangles: WS3.3-R; wine open stars: Sheng.}
	\label{fig:excluded_chains}
\end{figure*}

To test $\mathrm{BECR}_{1p}$ on unfitted data, Fig.~\ref{fig:excluded_chains} shows nine chains excluded from the 893-nucleus
fitting set: Tb, Tm, Lu, Re, Po, Rn, Fr, Ra, and Cm. The predictions use only the globally determined parameters, so this is a genuine out-of-fit test. Predictions from Sheng~\cite{Sheng2015}, WS3.3-R~\cite{WangLi2013}, and HFB-29~\cite{Goriely2016} are also shown. $\mathrm{BECR}_{1p}$ reproduces the general trend for these chains, comparable to the other models.

However, several chains show systematic offsets between experiment and all models. These are coherent shifts, not random scatter, and appear in WS3.3-R and HFB-29 as well. The offsets likely have multiple origins: on the experimental side, reference-radius normalization ambiguities in some chains~\cite{Angeli2013}; on the theoretical side, complex structures (shape coexistence, octupole deformation, pairing changes) in the rare-earth and actinide regions that any global formula struggles with. The RMSD for these nine chains is about $0.04$ fm, much larger than the 0.0138 fm for the fitted set. The overall trends are well reproduced, but the absolute offsets remain, pointing to the need for better deformation treatment or higher-order corrections, and for new measurements to resolve experimental normalization issues.

Compared to HFB-29 in Figs.~\ref{fig:isotopes_benchmark} and~\ref{fig:excluded_chains}, $\mathrm{BECR}_{1p}$ gives a lower global RMSD and often better local trends, especially in deformed and neutron-rich regions. This shows that $\epsilon$ indeed carries useful structural information. However, HFB-29, rooted in a well-defined energy density functional, offers a firmer foundation for extrapolation to drip-line nuclei and for predicting quantities beyond the empirical level. For exotic nuclei far from stability, microscopic models remain a more physically robust guide.

\subsection{Residual correction with AKRR and its applicability}
The BECR$_{1p}$ residuals still contain systematic information from higher-order shell effects, deformation coupling, or shape coexistence. To probe this, we apply an anisotropic kernel ridge regression (AKRR)~\cite{Wuxh24} as a diagnostic correction on top of BECR$_{1p}$. AKRR is not part of the formula; it is only a residual analyser.
Define $\Delta r = r_{\mathrm{ch}}^{\mathrm{expt}} - r_{\mathrm{ch}}^{\mathrm{th}}$, with $r_{\mathrm{ch}}^{\mathrm{th}}$ from BECR$_{1p}$. Using $(N,Z)$ as inputs, we optimize hyperparameters by LOOCV over the 893 points: $\sigma_1 = 4.6$, $\sigma_2 = 0.8$, $\lambda = 0.3$. The anisotropy ($\sigma_1 > \sigma_2$) means residual correlations extend more along $N$ than $Z$, reflecting smoother changes with neutron number and sharper shell effects with proton number. LOOCV evaluates the correction: for each nucleus, AKRR is trained on the other 892 residuals and predicts the held-out one; the corrected radius is $r_{\mathrm{ch}}^{\mathrm{AKRR}} = r_{\mathrm{ch}}^{\mathrm{th}} + \Delta r_{\mathrm{AKRR}}$, where $\Delta r_{\mathrm{AKRR}}$ is the LOOCV-predicted correction.

\begin{figure*}[htbp]
	\centering
	\includegraphics[width=0.95\textwidth]{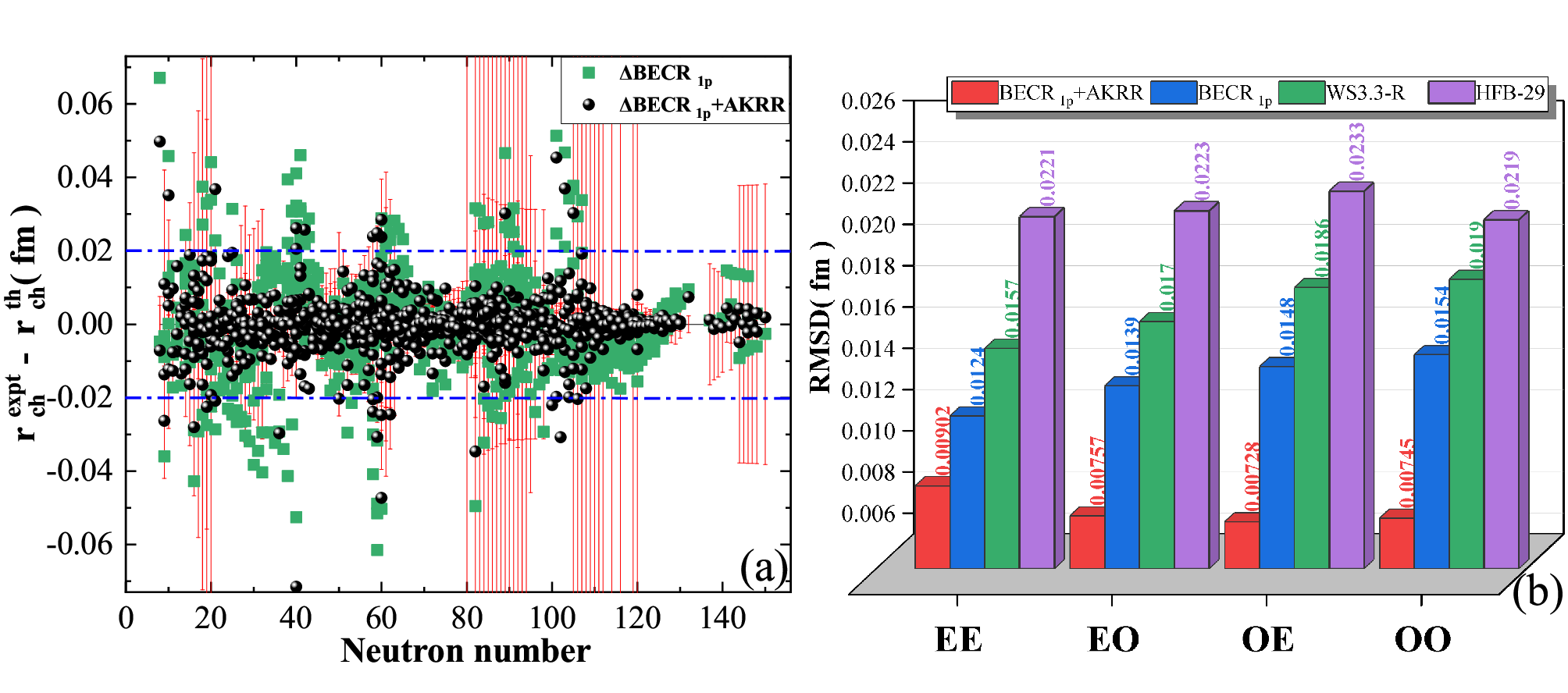}
	\caption{(Color online) Effect of the AKRR residual correction.
		(a) Residual distributions for 893 charge radii before (green squares) and after (black circles) AKRR correction. Red capped lines: experimental uncertainties; blue dash-dotted lines: $\pm0.02~\mathrm{fm}$ reference interval.
		(b) RMSD comparison for AKRR-corrected results, $\mathrm{BECR}_\mathrm{1p}$, WS3.3-R, and HFB-29 across odd-even categories.	}
	\label{fig:akrr_summary}
\end{figure*}

Figure~\ref{fig:akrr_summary} shows the effect of this
correction. The BECR$_{1p}$ residuals, already near zero, still show local fluctuations. After AKRR, the residual distribution is further compressed: the LOOCV RMSD drops from 0.0138 to 0.0081 fm, and 98.4\% of points lie within $\pm 0.02$ fm. Figure~\ref{fig:akrr_summary}(b) shows that all four odd-even categories benefit equally, indicating learnable systematic residuals across pairing types.

This $0.0081~\mathrm{fm}$ is an \emph{interpolation} accuracy within the 893-nucleus training region; the AKRR kernel acts as a local smoother. It is not the intrinsic precision of BECR$_{1p}$. For extrapolation to neutron-rich or proton-rich nuclei far from training data, AKRR becomes unreliable because it depends on nearby training coverage and on the quality of $\epsilon$ and deformation inputs. Thus, while AKRR-corrected values are useful near known regions, the predictions for the 11205 nuclei in the supplementary material should be used with caution in sparse-data areas; there, the physics-based BECR$_{1p}$ result is a safer baseline. For unmeasured nuclei, $\epsilon$ is taken from the WS3.3 mass table~\cite{WangWS2010}.

\section{Summary}
\label{sec:conclusion}
We have constructed a charge-radius formula that extracts the leading scale from $\epsilon = B/A$ and adds corrections for shell, odd-even, finite-size, and deformation effects. On 893 experimental radii, the RMSD drops from $0.0345$ fm (macroscopic BECR$^{\mathrm{mac}}$) to $0.0138$ fm (full formula). Parameter correlations reduce the formula to a single parameter, $\kappa = 0.7190$, with negligible loss of accuracy. The resulting BECR$_{1p}$ reproduces global trends and local features---odd-even staggering, shape-transition jumps, shell-closure kinks---across isotopic chains; the nine unfitted chains reveal the limits of any global empirical systematics. With only one parameter, BECR$_{1p}$ does not replace microscopic models but offers a complementary, physically transparent view of the binding--radius correlation. As a residual diagnostic, AKRR lowers the LOOCV RMSD to about $0.0081$ fm, showing that systematic information remains in the residuals and can be recovered data-dependently. Predictions for 11205 nuclei are provided as supplemental material~\cite{Tianjl26}.

\begin{center}
\textbf{ACKNOWLEDGMENTS}
\end{center}
This work was supported by the Guangxi Science and Technology Program (No. 2023GXNSFDA026005 and No. 2023GXNSFBA026008), the National Natural Science Foundation of China (No. 12465019 and No. 12465021), and the Central Government Guides Local Scientific and Technological Development Fund Projects (No. Guike ZY22096024).

\end{document}